\documentclass[fleqn,usenatbib]{mnras}

\usepackage{caption}
\usepackage{newtxtext,newtxmath}
\usepackage{subfigure}
\usepackage{tabularx}
\usepackage[T1]{fontenc}

\DeclareRobustCommand{\VAN}[3]{#2}
\let\VANthebibliography\thebibliography
\def\thebibliography{\DeclareRobustCommand{\VAN}[3]{##3}\VANthebibliography}

\usepackage{graphicx}	% Including figure files
\usepackage{amsmath}	% Advanced maths commands
\title[Terminator Region Atmosphere of WASP-77Ab]{The Terminator Region Atmosphere of the hot Jupiter WASP-77Ab with ESPRESSO/VLT observations}

\author[Zewen Jiang et al.]
{
Zewen Jiang$^{1,2}$,
Wei Wang$^{1}$\thanks{E-mail: wangw@nao.cas.cn},
Guo Chen$^{3}$,
Yaqing Shi$^{4}$,
Meng Zhai$^{5}$,
Patricio Rojo$^{6}$,
Yujuan Liu$^{1}$,
Gang Zhao$^{1,7}$
\\
$^{1}$CAS Key Laboratory of Optical Astronomy, National Astronomical Observatories, Chinese Academy of Sciences, Datun Road A20, Beijing 100101, China\\
$^{2}$CAS Key Laboratory of Particle Astrophysics, Institute of High Energy Physics, Chinese Academy of Sciences, Beijing 100049, China\\
$^{3}$CAS Key Laboratory of Planetary Sciences, Purple Mountain Observatory, Chinese Academy of Sciences, Nanjing 210023, China\\
$^{4}$Space Engineering University, Beijing 101416, China\\
$^{5}$CAS South America Center for Astronomy, National Astronomical Observatories, Chinese Academy of Sciences, Datun Road A20, Beijing 100101, China\\
$^{6}$Departamento de Astronomía, Universidad de Chile, Camino El Observatorio 1515, Las Condes, Santiago, Chile\\
$^{7}$School of Astronomy and Space Science, University of Chinese Academy of Sciences, Beijing 100049, China\\
}
\date{Accepted 2024 December 01. Received 2024 December 01; in original form 2023 November 07}

\pubyear{2024}

\begin{document}
\label{firstpage}
\pagerange{\pageref{firstpage}--\pageref{lastpage}}
\maketitle

\begin{abstract}
Atmospheric studies are essential for elucidating the formation history, evolutionary processes, and atmospheric dynamics of exoplanets. High-resolution transmission spectroscopy offers the advantage of detecting subtle variations in stellar spectral profiles, thereby enabling the identification of the sources of observed signals. In this study, we present the transmission spectra of the exoplanet WASP-77A\,b, a hot Jupiter with a 1.36-day orbital period around a G8 host star with  $V=11.29$\,mag. These observations were conducted using the high-resolution spectrograph ESPRESSO at the Very Large Telescope over three transit events. We analyze the Rossiter-McLaughlin effect for WASP-77A and determine a projected spin-orbit angle of ${\lambda = 16.131^{\circ}}^{+2.106}_{-2.324}$, indicating that the planet's orbit is nearly aligned. Following the generation of transmission spectra for the three nights, we model and correct for center-to-limb variation and the Rossiter-McLaughlin effects. In the residual transmission spectra, we detect H$\alpha$, H$\beta$ and \ion{Ca}{ii} H with a significance exceeding 3.5$\sigma$. After applying $0.1-0.5$\AA\ masks to the cores of these lines to mitigate stellar contamination, all them still shows visible absorptions although not significant, suggesting at least partial planet contribution to them. Therefore, we are yet unable to confirm or reject the planetary origin of these spectral signals based on the current data set. Further investigation of WASP-77Ab's atmosphere, particularly in areas beyond the terminator region, is essential to illuminate the planet's two-dimensional atmospheric structure.

\end{abstract}


%%%%%%%%%%%%%%%%%%%%%%%%%%%%%%%%%%%%%%%%%%%%%%%%%%

%%%%%%%%%%%%%%%%% BODY OF PAPER %%%%%%%%%%%%%%%%%%

\section{Introduction} 
\label{sec:intro}
In recent decades, Hot Jupiters (HJs) have emerged as one of the most significant discoveries in the field of astronomy, greatly enhancing our understanding of distant and previously uncharted worlds~\citep{Mayor_1995}. Characterized by their elevated equilibrium temperatures, low occurrence rates of approximately 1\%, and distinctive orbital properties, HJs offer a unique opportunity to examine the atmospheric compositions and investigate the temperature-pressure (T–P) profiles of gaseous planets subjected to intense irradiation ~\citep{Marcy_2005,Cumming_2008,Mayor_2011,WangJi_2015}.

Many atoms and ions have been reported in the atmospheres of HJs, such as Na, K,  Mg, Ca, V, Co, Ti, Fe, Co+, Ca+, Fe+, Ti+, V+, Sc+, Sr+, Ni+, Ba+, Rb and Sm ~\citep{Snellen_2010,Wyttenbach_2015,Hoeijmakers_2018,Yan_2018,Casasayas_Barris_2019,Chen_2020,Seidel_2020,Dorval_2020,Ehrenreich_2020,Tabernero_2021,Kesseli_2022,Lira-Barria_2022, Silva_2022, Jiang_2023a}. These species are detected using the transmission spectroscopy observed by the ground-based high-resolution spectrographs. As one of the most powerful tools for characterizing the exoplanet atmosphere, high-resolution spectroscopy (HRS) can separate the planet lines well from the stellar lines due to their different radial velocities, and thus determine whether the detected signal is from the planet or from the Rossiter-McLaughlin effect~\citep{Rossiter_1924,Ohta_2005}. For the same reason, HRS can provide information on the relative motions of different species, and thus can also be used to study the dynamics of planet atmospheres.

WASP-77A\,b is a classic HJ, possessing a mass of $1.76\pm0.06$\,$M_{\rm Jup}$, a radius of $1.21\pm0.02$\,$R_{\rm Jup}$, an equilibrium temperature $T_{\rm eq}$ of $1715^{+26}_{-25}$\,K and an orbital period of $\sim$1.36\,d~\citep{Maxted_2013,Cort_2020}. The host star, WASP-77A, is accompanied by a K-dwarf star, designated WASP-77\,B, which exhibits a $\sim$2.2\,mag difference in $V$ and is located $\sim3.3$\arcsec away~\citep{Maxted_2013}. The discovery of  WASP-77A\,b~\citep{Maxted_2013} was achieved through photometric observations conducted by the ground-based Wide Angle Search for Planets (WASP)~\citep{Pollacco_2006} and radial velocity (RV) measurements utilizing the fibre-fed CORALIE spectrograph~\citep{Queloz_2000_CORALIE} on the Euler 1.2 m telescope. Further confirmation was provided by photometric data collected with the 60\,cm TRAnsiting Planets and PlanetesImals Small Telescope (TRAPPIST)~\citep{Jehin_2011} and RV measurements obtained from the High Accuracy Radial velocity Planet Searcher (HARPS) ~\citep{Mayor_2003} spectrograph at the ESO 3.6\,m telescope. Additionally, subsequent photometric observations were conducted using the three channel photometer ULTRACAM~\citep{Dhillon_2007}, mounted on the 4.2\,m William Herschel Telescope (WHT), to mitigate the influence of nearby stars, thereby confirming that the small object was indeed transiting the bright star. Given its relatively high equilibrium temperature, WASP-77A\,b presents a compelling target for atmospheric characterization studies. 

Recently, \citet{Line_2021} conducted the first detection of H$_2$O and CO in the dayside atmosphere of WASP-77A\,b utilizing the Immersion GRating INfrared Spectrometer (IGRINS) at the Gemini South Observatory. They quantified the atmospheric gas volume mixing ratios and observed a solar carbon-to-oxygen (C/O) ratio alongside sub-solar metallicity. Their findings led to the conclusion that the scenarios of disk-free migration and a carbon-rich atmosphere are unlikely applicable to WASP-77A\,b. Subsequently, \citet{Mansfield_2022} corroborated the presence of H$_2$O absorption in the thermal emission spectrum of WASP-77A\,b through observations made with the Wide Field Camera 3 (WFC3) on the Hubble Space Telescope (HST). More recently, \citet{August_2023} reaffirmed the detection of thermal emission signals for H$_2$O and CO in WASP-77A\,b, utilizing the NIRSpec instrument aboard the James Webb Space Telescope (JWST), and reported results consistent with the metallicity and C/O ratio findings of \citet{Line_2021}. However, to date, there have been no studies reported concerning the terminator region of WASP-77A\,b.

This paper is organized as follows. Section~\ref{sec:observations and data reduction} provides a comprehensive overview of the three ESPRESSO transit observations conducted on WASP-77A\,b, along with the associated data reduction processes. The methodologies employed for data analysis, which encompass the generation of the transmission spectrum and the atmospheric cross-correlation analysis, are delineated in Section~\ref{sec:DATA ANALYSIS}. Finally, Section~\ref{sec:Searching for atmospheric species} presents the findings and subsequent discussions.

\section{OBSERVATIONS AND DATA REDUCTION}
\label{sec:observations and data reduction}
\begin{figure}
   \centering
   \includegraphics[width=8cm, height=12cm]{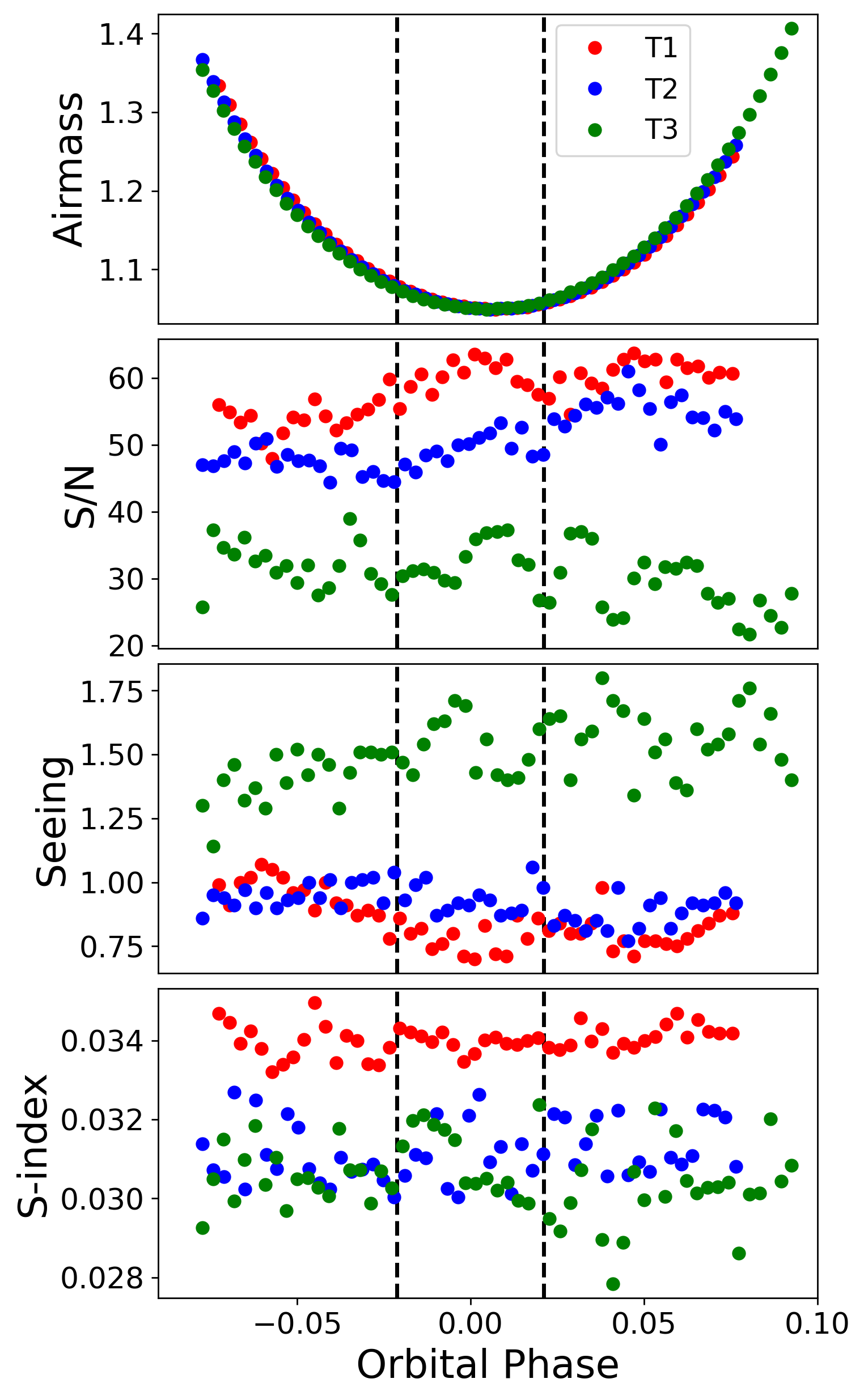}
 \caption{The temporal variations of the airmass, SNR@550\,nm, seeing and S-index of the three transit observations from top to bottom, respectively. The black dashed lines indicate the first and fourth contacts of the transits.}
    \label{fig:obs_conditions}
    \end{figure}

We conducted observations of three transits of WASP-77Ab on October 29, and November 13 and 28 in 2021, designated as T1, T2, and T3, respectively. These observations were carried out using the Echelle SPectrograph for Rocky Exoplanets and Stable Spectroscopic Observations (ESPRESSO), which is installed on the 8.2-meter Very Large Telescope (VLT) at the European Southern Observatory in Cerro Paranal, Chile \citep{Pepe_2021}.  ESPRESSO operates within the optical wavelength range of 380-788\,nm and achieves a spectral resolution of approximately R$\sim$140,000 in high-resolution mode. All observations were conducted under the program 0108.C-0148(A) (Principal Investigator: Wei Wang), utilizing the HR21 observing mode. In total, we acquired 157 exposures, each lasting 300 seconds, which included 22 in-transit spectra and 27 out-of-transit spectra for T1. These observations encompassed the planet's orbital phase $\phi$ from $-0.072$ to $+0.075$, yielding an average signal-to-noise ratio (SNR) of 58 near 550\,nm. For T2, we obtained 23 in-transit spectra and 28 out-of-transit spectra, with an average SNR of 51, covering the orbital phase $\phi$ from $-0.077$ to $+0.076$. For T3, we collected 22 in-transit spectra and 35 out-of-transit spectra, achieving an average SNR of 31 with $\phi$ varying from $-0.077$ to $+0.092$. A comprehensive summary of the observational details is provided in Table~\ref{obs}. The temporal variations of airmass, SNR, seeing, and S-index during the three observing nights are illustrated in Fig.~\ref{fig:obs_conditions}, clearly depicting the observational conditions for each transit. Standard echelle data reduction procedures were applied to the raw 2D spectra using the ESPRESSO reduction pipeline (version 2.3.4), which included corrections for bias, dark current, flat-fielding, spectral extraction and wavelength calibration. The final wavelength-calibrated extracted 1D spectra contain information on the vacuum frame wavelengths, stellar fluxes, and associated uncertainties, which serve as the basis for our subsequent analysis.

It is observed that the companion star WASP-77B, which is 2.2 magnitudes fainter, is located at an angular separation of $\sim3.3$\arcsec from WASP-77A, as reported by \cite{Maxted_2013}. This separation exceeds the size of the ESPRESSO fiber of 1\arcsec. Utilizing a two-dimensional Gaussian point spread function (PSF) with seeing conditions of 1\arcsec and 1.8\arcsec, the flux contamination from WASP-77B to WASP-77A is estimated to be $<0.0001$\% and less than 0.0043\%, respectively. This indicates that, particularly for the T1 and T2 datasets, which were predominantly acquired under seeing conditions better than 1.0\arcsec, the influence of contamination from WASP-77B is completely negligible. This contamination is also considered almost innocent for the T3 data.

%------------------------------Airmss-SNR-------------------------%
%----------------------------------------------------------------%

%Another one transit of WASP-77Ab was observed on 06 Nov 2018, using High Accuracy Radial velocity Planet Searcher (HARPS) echelle spectrograph on the ESO 3.6m telescope, La Silla, in chile (Mayor et al. 2003) which covers the optical range between 380nm and 690nm with a spectral resolution of $R\sim115000$. Data were retrieved from the ESO archive under the programs 0102.C-0618(A) (PI:ESPOSITO). 21 spectra in total were obtained, including 9 in-transit spectra and 12 out-of-transit spectra with an average S/N of 48 covering the phase $\Phi$ from -0.078 to +0.08. The integration time of each exposure is about 900s. The data were reduced with version 3.8 of the HARPS Data Reduction Pipeline and the wavelength item are given in air.  Bias, dark, flat and bad pixels were taken into account when processing the 2D images, from which 1D spectra can be sky-subtracted and extracted, then all the odrers data were wavelength calibrated and stitched into 1d spectra. The related observed information of data sets are described in Table~\ref{obs}.

%%%%%%%%%%%%%%%%%%%%%%%%%%%%%%%%%%%%%%%%%%%%%%%%%%%%%
   \begin{table*}
      \centering
      \caption{Summary of the WASP-77Ab transit observations.}
         \label{obs}
         \begin{tabular}{lcccccccc}
            \hline
            \hline
            \noalign{\smallskip}
            & Date &\multicolumn{3}{c}{Number of spectra} & Exp. time & Airmass range & Mean SNR & Program ID\\
            \cline{3-5}
            & (UT) &  Total & In-transit & Out-of-transit & (s) &(start$-$T$_{\rm c}-$end)  & (@550\,nm) &   \\
            \hline
            T1 & 2021-10-29 & 49 & 22 & 27 &300 & $1.99-1.05-1.91$ & $\sim$58 & 0108.C-0148  \\
            T2 & 2021-11-13 & 51 & 23 & 28 &300 & $1.72-1.05-2.46$ & $\sim$51 & 0108.C-0148 \\
            T3 & 2021-11-28 & 57 & 22 & 35 &300 & $1.72-1.05-2.46$ & $\sim$31 & 0108.C-0148 \\
            %HARPS & 2018-11-06 & 21 & 9 & 12 &900 & $1.72-0-2.46$ & $\sim$48 & 0102.C-0618 \\
            \hline
         \end{tabular}
         \newline
     \end{table*}
%%%%%%%%%%%%%%%%%%%%%%%%%%%%%%%%%%%%%%%%%%%%%%%%%%%%%

%%%%%%%%%%%%%%%%%%%%%%%%%%%%%%%%%%%%%%%%%%%%%
% First figure
\begin{figure}
	\includegraphics[width=\columnwidth]{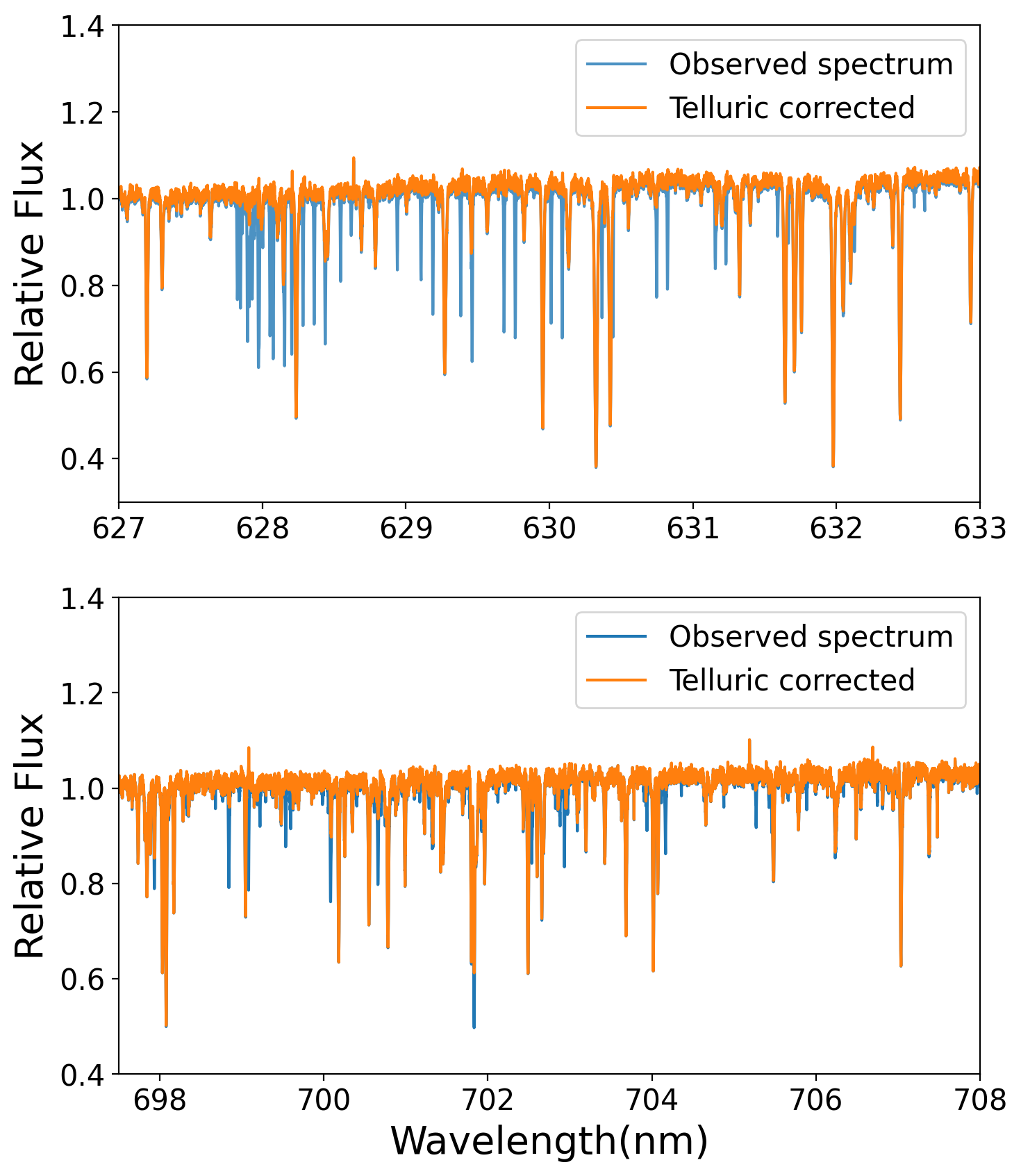}
    \caption{An example of telluric correction using \texttt{Molecfit} in the observed spectrum of WASP-77A\,b taken in T1 around 630\,nm (\textit{Top}) and 703\,nm(\textit{Bottom}). The observed spectra are shown in blue, while the spectra after removing the telluric contamination are shown in orange. The dashed green vertical lines represent the static positions of the atomic lines in vacuum.}
    \label{fig:telluric_correction_figure}
\end{figure}
   \begin{table}
   %\centering
   
      \caption{Parameters of the WASP-77A system.}
         \label{tab:system} 
           \begin{tabular}{lll}
            \hline
            \hline
            Description &Symbol & Value \\
            \hline
            Stellar Parameters & & \\
            \hline
            $V$ magnitude& $m_{\rm v}$    & $10.30\pm0.05$\,mag    \\
            Effective temp& $T_{\rm eff}$& $5500\pm80$\,K   \\          
            Surface gravity&  log\,$g_\star$  & $4.33\pm0.08$ cgs\\    
            Metallicity& [Fe/H] &  $0.00\pm0.11$\,dex   \\
            Stellar mass&  $M_\star$  & $1.002\pm0.045\,M_\odot$   \\
            Stellar radius &$R_\star$ & $0.955\pm0.015\,R_\odot$   \\
            Projected spin&V\rm sin$i_{\star}$ & $4.0\pm0.02\,\rm {km\,s^{-1}}$\\
            \noalign{\smallskip}
            \hline
		    Planet Parameter & & \\
            \hline
            \noalign{\smallskip}
            Planet mass  & $M_{\rm p}$ & $1.76\pm0.06\,M_{\rm Jup}$   \\  
            Planet radius & $R_{\rm p}$ & $1.21\pm0.02\,R_{\rm Jup}$ \\           
            Planet density$^{a}$ &$\rho$  & $1.33\pm0.04$\,g\,cm$^{-3}$ \\          
            Equilibrium temp$^{a}$& $T_{\rm eq}$ & $1715_{-25}^{+26}$\,K  \\  
            Radius ratio$^{a}$& $R_{\rm p}/R_\star$ & $0.13354_{-0.00070}^{+0.00074}$\\
            \noalign{\smallskip}
            \hline
            Orbit Parameters& &\\
            \hline
            Epoch-BJD& $T_{\rm c}$ & $2455870.44977\pm0.00014$ \\ 
            Period & $P$  & $1.3600309\pm0.000002$\,d    \\ 
            Transit duration& $T_{14}$ & $0.09\pm0.00035$\,d \\ 
            Semi-major axis & $a$ &  $0.024\pm0.00036$\,AU   \\ 
            Inclination &$i$ & $89.4_{-0.7}^{+0.4}$\,deg \\   
            Semi-amplitude velocity$^{a}$&$K_{\star}$&$323.4_{-3.4}^{+3.8}$\,\rm m$\,\rm s^{-1}$\\
            \hline
            \noalign{\smallskip}
         \end{tabular}
        \textbf{Notes}:$(a)$~\citet{Cort_2020}. All other parameters refer to~\citet{Maxted_2013}.
   \end{table}
%%%%%%%%%%%%%%%%%%%%%%%%%%%%%%%%%%%%%%%%%%%%%%%%%%

\section{DATA ANALYSIS}
\label{sec:DATA ANALYSIS}
\subsection{Telluric correction}
\label{sec:Telluric correction}
Following the earlier works conducted by \cite{Allart_2017} and \cite{Jiang_2023a}, we utilized the ESO software \texttt{Molecfit}, version 1.5.7~\citep{Smette_2015,Kausch_2015}, to mitigate the additional absorption lines introduced by Earth's atmosphere. \texttt{Molecfit} is a sophisticated tool that incorporates a publicly available radiative transfer code, a molecular line database, and atmospheric profiles, enabling the correction of telluric absorption lines through synthetic modeling of the Earth's atmospheric transmission~\citep{Smette_2015}. This software has gained considerable traction in recent high-resolution spectroscopy (HRS) studies of exoplanet atmospheres \citep[e.g.,][]{Allart_2017, Alonso-Floriano_2019, Kirk_2020}, facilitating the effective removal of telluric contamination.

The input parameters for \texttt{Molecfit} were selected in accordance with the methodologies outlined in \citet{Allart_2017, Jiang_2023a, Jiang_2023b}. To enhance the efficiency and accuracy of the telluric contamination removal process, we manually defined the fitting region by comparing the observed spectra with the model telluric atmospheric transmission spectra. The ESPRESSO spectra, acquired from the ESO pipeline, are initially presented in the solar system barycentric rest frame and subsequently shifted to the terrestrial rest frame to comply with the requirements of \texttt{Molecfit} prior to the application of the telluric correction. In Fig.~\ref{fig:telluric_correction_figure}, we illustrate the comparison between the observed spectrum obtained on T1 and the corresponding telluric contamination-corrected spectrum in the vicinity of the 630\,nm and 700\,nm in the top and bottom panels, thereby demonstrating the efficacy of our telluric correction approach.

\subsection{Rossiter-McLaughlin analysis}
\label{sec:rm analysis}
Due to the stellar rotation, the RV curve may deviates slightly from a Keplerian orbit during the transit of a planet across the stellar disk. This phenomenon occurs due to variations in projected velocities and brightness characteristics, leading to additional signals that are superimposed on the Doppler reflex motion. This phenomenon, known as the RM effect~\citep{Ohta_2005}, was initially identified by \citet{Rossiter_1924} and \citet{McLaughlin_1924} in the context of binary stars. The same effect in a planetary system was first observed by \citet{Queloz_2000}. Observing this effect can place constraints on the impact parameters of the transit and measured the product of the stellar rotational velocity and the sine of the inclination angle (\( \upsilon \sin i_{\star} \)), as well as the angle between the orbital plane and the apparent equatorial plane, commonly referred to as the projected spin-orbit angle (\( \lambda \)), through modeling of the residual RV curves. In the present study, the RV values and their associated uncertainties derived from the ESPRESSO pipeline utilizing the cross-correlation function (CCF) method are employed.

In order to analyze and model the RM effect, we employed the Markov Chain Monte Carlo (MCMC) algorithm implemented in the \texttt{emcee} package~\citep{Foreman-Mackey_2013}. We utilized the RM effect model available in PyAstronomy~\citep{Czesla_2019}, following the methodology outlined in \citet{Ohta_2005}, which provides an analytical fit for the RM effect associated with a circular orbit. The parameters considered as free variables in our analysis included the spin-orbit angle ($\lambda$), the linear limb-darkening coefficient ($\epsilon$), the stellar angular rotation velocity ($\Omega$), and the inclination of the stellar rotation axis ($i_\star$). The remaining parameters, namely $a$, $P$, $i$, $\textit{T}_c$, and the ratio of the planetary radius to the stellar radius ($R_{p}/R_{\star}$), were held constant at values reported in the literature, as detailed in Table~\ref{tab:system}. 

\begin{figure}
   \centering
   \includegraphics[width=8cm, height=9cm]{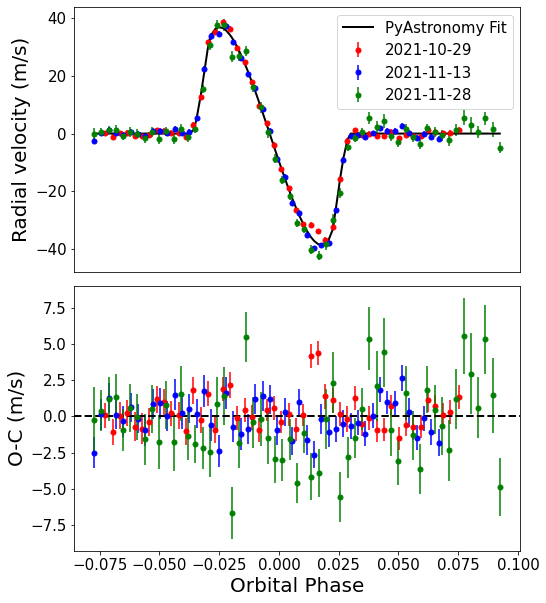}
 \caption{{\textit{Top panel}:} The RV curve with Kepler motion removed, showing the RM effect of WASP-77Ab of the three transits represented by the red, green, and blue solid circles with error bars. The best-fit model using the PyAstronomy package is shown in the solid line. \textit{Bottom panel}: The residuals after removing the model prediction from the data.}
    \label{fig:RM fit}
    \end{figure}

To facilitate a comprehensive examination of the distribution of free parameters, we employed a configuration of 100 walkers, each executing 50,000 steps. The initial 10,000 steps were excluded as burn-in, allowing for an effective exploration of the parameter space and enabling the probability density to stabilize within the region of maximum likelihood in the MCMC analysis. The chosen priors and the resulting posteriors for each parameter derived from the MCMC fitting are presented in Table~\ref{prior}. Additionally, the distribution of the calculated posterior using data from three nights is illustrated in Fig~\ref{fig:RM_fit_corner}. The median values of the posterior distributions for the parameters $\lambda$ and $\epsilon$ are 16.131$^{\circ}$ $^{+2.106}_{-2.324}$ and $0.893^{+0.013}_{-0.013}$, respectively.
Using the archived HARPS and HARPS-N data \citet{Zak_2024} determined a $\lambda$ of -8$^{\circ}$$^{+19}_{-18}$, which differs slightly from our determined $\lambda$ by $\sim 1.1\sigma$. Nevertheless, both values are relatively small, suggesting that the planetary orbital plane is approximately aligned with the stellar equator. Given that our ESPRESSO spectra have higher SNRs compared to the HARPS and HARPS-N data, we used our determined $\lambda$ for the further analysis. The median value of $v{\rm sin}\,i_{\star}$ is $3.28^{+0.040}_{-0.035}\,\rm km\,s^{-1}$, which was determined by taking into account $\Omega$ and $i_\star$. These median values are utilized as the best-fit parameters for modeling the RM effect. The phase-folded RV curve alongside the best-fit model is depicted in Fig~\ref{fig:RM fit}, accompanied by a residual plot that exhibits a root mean square (rms) residual of approximately 1.923\,m\, s$^{-1}$. We employed these values as input parameters to model and correct for the center-to-limb variation (CLV) effect, which arises from the nonuniform brightness across the stellar disk, as well as the RM effect resulting from variations in projected RVs at different distances from the stellar center, in order to mitigate their impact on the derived transmission spectra.

\subsection{Construction of Planet Transmission Spectra}
\label{sec:Transmission spectroscopy}
In order to investigate the presence of atmospheric species on exoplanets, we employed the methodology established by \citet{Wyttenbach_2015} and \citet{Casasayas-Barris_2019} to derive transmission spectra for each observational exposure. Following the removal of telluric absorption as detailed in Section~\ref{sec:Telluric correction}, we utilized the \texttt{iSpec} package~\citep{Blanco-Cuaresma_2014,Blanco-Cuaresma_2019} to normalize all observed spectra to their respective continuum levels. Additionally, we referenced the work of \citet{Allart_2017} to implement a sigma-clipping rejection algorithm on the normalized spectra, thereby substituting anomalous values resulting from cosmic ray impacts with the mean value of the other spectra at each pixel. To ensure alignment of the stellar lines, we adjusted all spectra to their stellar rest frame, taking into account barycentric Earth radial velocity (BERV), $V_{\rm sys}$, and the stellar reflex motion induced by the planet, $V_{\rm reflex}$. Specifically, BERV is obtained directly from the files header information of the ESPRESSO spectra, while $V_{\rm sys}$ is derived from the RM analysis (as described in Sect.~\ref{sec:rm analysis}) and $V_{\rm reflex}$ is calculated using the orbital parameters as listed in Table~\ref{tab:system}). 

Subsequently, we constructed a master-out spectrum, which is expected to be of purely stellar origin, by averaging all out-of-transit spectra, weighted by their mean SNRs. This master-out spectrum was then used to divide each individual spectrum, effectively eliminating the contribution of stellar spectra and allowing the absorption signal from the planetary atmosphere to be more prominent~\citep{Stangret_2021}. Then, each divided in-transit spectrum is shifted to the planetary rest frame (PRF) by incorporating the planet's RV, denoted as $V_{\rm p}$. This value is determined from the Kelpler's orbital velocity $K_{\rm p}$ and the orbital phase at a specific time, while $K_{\rm p}$ is calculated using the orbital parameters $a$ and $P$ provided in Table~\ref{tab:system} assuming a circular orbit. By aggregating the shifted residual spectra, a 2D transmission spectrum map is produced, which is subsequently utilized for further CCF analysis (cf. details in Section~\ref{sec:ccf_analysis})

Finally, the integrated 1D planet transmission spectrum is obtained by summing up the S/N-weighted in-transit PRF residuals, using the following formula:
%------------------------------------------
\begin{equation} \label{prv}
        \Re(\lambda) = \sum_{\rm in}{\frac{F_{\rm in}(\lambda)}{F_{\rm out}}}\bigg|_\textnormal{{Planet\  RV\ Shift}}\      -1
,\end{equation}
%-----------------------------------------

%---------------------------------------------------
%-----------------------------------------------------

%---------------------RM effect fit prior table--------
   \begin{table*}
    
      \caption{Parameters derived from the RV curve fitting.}
         \label{prior}
         
         \begin{tabular}{p{0.28\linewidth}cccc}
            \hline
            \hline
            \noalign{\smallskip}
             Description &Symbol& Prior & Posterior   \\
            \noalign{\smallskip}
            \hline
            \noalign{\smallskip}
           Projected spin-orbit angle&$\lambda$ &U(-50,\,50)&$16.131^{+2.106}_{-2.324}$\,deg& \\
            \noalign{\smallskip}
           Linear limb dark coefficient&$\epsilon$ & U(0.5,\,1)&$0.893^{+0.013}_{-0.013}$\\
            \noalign{\smallskip}
           Projected stellar rotation velocity&$\Omega$ &U($1\rm e^{-6}$,\,$1\rm e^{-5}$)&$0.000004997^{+1.06\rm e^{-7}}_{-7.30\rm e^{-8}}$ rad\,s$^{-1}$ \\  
            \noalign{\smallskip}
           Inclination of stellar rotation axis&$i_\star$ & U(50,\,150) & $87.999^{+10.570}_{-9.500}$\,deg \\  
            \noalign{\smallskip}
            \hline
            \noalign{\smallskip}
         \end{tabular}
         
         \textbf{Notes}:\textit{U(a,b)} represents a uniform distribution with a low and high limit of a and b, respectively.
   \end{table*}
   
%----------------------------------------------
\subsection{Removal of the RM and CLV Effect}
\label{sec:remove rm and clv effect}

The CLV and RM effects can significantly alter the stellar line profile, potentially introducing erroneous signals into the final transmission spectra of planetary atmospheres. To accurately extract the planetary signal, it is essential to eliminate the CLV and RM effects from the two-dimensional transmission spectrum maps, as outlined in Section~\ref{sec:Transmission spectroscopy}, as well as from the CCF maps discussed in Section~\ref{sec:ccf_analysis}. In this study, we adopt the methodology utilized by \citet{Yan_2018}, \citet{Chen_2020}, and \citet{Casasayas_Barris_2022} to model the stellar spectra at various transit positions. We employ the \texttt{Spectroscopy Made Easy} (SME) tool~\citep{Valenti_1996} to compute theoretical stellar spectra at 21 distinct limb-darkening angles ($\mu$), utilizing the MARCS and VALD3 line lists~\citep{Ryabchikova_2015}, while assuming solar abundance and local thermodynamic equilibrium (LTE) for the spectral calculations. The stellar disk is partitioned into small elements measuring $0.01\,R_\star \times 0.01\,R_\star$, each characterized by specific parameters that account for the projected rotational velocity ($v\,{\rm sin}\, i_{\star}$), limb-darkening angle ($\mu$), and the angle ($\theta$) between the normal to each small element and the line of sight. The relative position of the planet with respect to its host star is determined from the orbital parameters at any given phase, under the assumption of uniform velocity during the transit. Subsequently, the synthetic spectrum is generated by aggregating the spectra from all elements, excluding those that are obscured by the planet.

The subsequent procedure involves simulating the RM+CLV effects and eliminating them. The ways of removing the RM+CLV effects for single lines and for the CCFs are different. Specifically, for each single line we re-scale the obtained transmission spectra around the line center to match with the RM+CLV effects models. While for the species with multiple lines, we perform a cross-correlation with the template spectra generated by \texttt{petitRADTRANS} for each species. The resulting 2D CCF map of RM+CLV effects serves as a proxy, as illustrated in  Figure~\ref{fig:RC_model_and_ccfexample}, which will be re-scaled to match with the CCF map calculated by observational data. To match the RM+CLV effects model with the CCF map, we implement Chi-Square minimization utilizing the MCMC algorithm as realized by \texttt{emcee}~\citep{Foreman-Mackey_2013}, selecting  parameters that correspond to the maximum probability. Following this correction, the influence of the RM and CLV effects should be significantly diminished. 
%\cc{For the correction of RM+CLV effects in single lines, it is sufficient to re-scale the transmission spectra around single lines to match with the RM+CLV effects model calculated by the planet transits before its stellar disk with rotation and nonuniform brightness.}
Figure~\ref{clvrm} presents the CLV and RM effects in the residual spectra around the Na D1 and D2 lines around single lines, demonstrating the RM+CLV effect fitting. It seems that, however, for the case of WASP-77Ab, such correction is not optimal yet. 

\section{Searching for atmospheric species}
\label{sec:Searching for atmospheric species}

\subsection{Strong atomic lines}
\label{secstronglines} 
We firstly conduct a visual inspection of the phase-resolved transmission spectra around each prominent feature, which aids in verifying the presence of the targeted atomic lines. Figure~\ref{fig:atom line} presents the 2D phase-resolved transmission spectra, along with the S/N-weighted and combined spectrum for the Na, H$\alpha$, H$\beta$, \ion{Ca}{II} H, and K lines. These spectral lines have undergone corrections for the RM+CLV effects, with the specific methodologies for mitigating contamination in the transmission spectra detailed in Section~\ref{sec:remove rm and clv effect}. Furthermore, we employed a Gaussian function to determine the the center, depth, and width of each line. The results of the Gaussian fitting are compiled in Table~\ref{tab:atomic_lines}. Additionally, this table includes the line depths corresponding to various atoms, both with and without the exclusion of the core regions of these strong lines using a 0.1\AA\,mask, as well as the observed Doppler shift of the line center ($V_{wind}$), which traces the planetary winds directed towards the observer from the morning hemisphere to the evening hemisphere. The effective wavelength-dependent planetary radius, $R_{\lambda}$, is derived using the formula $\sqrt{1+h/\delta}$~$R_p$, where $\delta$ represents the transit depth of the planet.

\subsection{Cross-correlation search}
\label{sec:ccf_analysis} 
The search of the species exhibiting multiple spectral lines can be prompted by employing the CCF technique, which aims to determine the optimal correspondence between the calculated template spectra and the acquired transmission spectra. In this study, we investigate the presence of several species including Cr, Fe, Ti, V, Y, TiO, VO, HCN, H$_2$O, CaH and FeH. The template spectra are generated using the python package \texttt{petitRADTRANS}~\citep{P_Molliere_2019}, which has been effectively utilized in numerous atmospheric studies~\citep{Casasayas_Barris_2022, Jiang_2023a,Jiang_2023b, Ouyang2023a, Ouyang2023b, Shi2023}. For input parameters, we consider the planet WASP-77Ab has a radius of 1.21$R_{\rm J}$ and a mass of 1.76$M_{\rm J}$, resulting in a log\,$g_{\rm p}$ of 3.441 cgs~\citep{Maxted_2013}. We employ an isothermal pressure-temperature (PT) profile with a temperature of 1700\,K and assume a solar abundance to ascertain the volume mixing ratio (VMR) of the various species. The template spectra produced by \texttt{petitRADTRANS} are convolved with a Gaussian function to align with the resolution of our ESPRESSO spectra. We utilize a RV range of $\pm200$\,km\,s$^{-1}$ with a step size of 0.5 km\,s$^{-1}$ and compute CCF for each residual spectrum in conjunction with the template spectra. The CCF is calculated using the following equation:
\begin{equation} \label{ccf}
\centering
        c(v,t)= \frac{\sum_i^N x_i(t) T_i(v)}{\sum_i^N T_i(v)},
\end{equation}

where $c(v,t)$ represents a 2D matrix that is dependent on both $t$ and $v$, $x_i(t)$ denotes the transmission spectrum at a specific $t$, while $T_i(v)$ refers to the template that has been shifted to a radial velocity of $v$. Should the species under investigation be present, a signal will manifest at the locations corresponding to the estimated orbital velocity$K_{\rm p}$ and the system velocity $V_{\rm sys}$\citep{Snellen_2010,Hoeijmakers_2019}.

The computed CCF is a 2D matrix that is a function of $t$ and $v$. In instances where there is a significant concentration of the targeted species within a planet's atmosphere, the corresponding signal will manifest as a trace along the anticipated trajectory of the planet within the CCF map. This signal will reach its peak at the coordinates corresponding to the estimated $K_{\rm p}$ and $V_{\rm sys}$ within the $K_{\rm p}-\Delta V_{\rm sys}$ map, as outlined by \citet{Hoeijmakers_2019}. To enhance the SNRs and improve detection significance, we aggregate the CCF maps obtained over three nights into a singular CCF map.

\begin{figure}
	\includegraphics[width=\columnwidth]{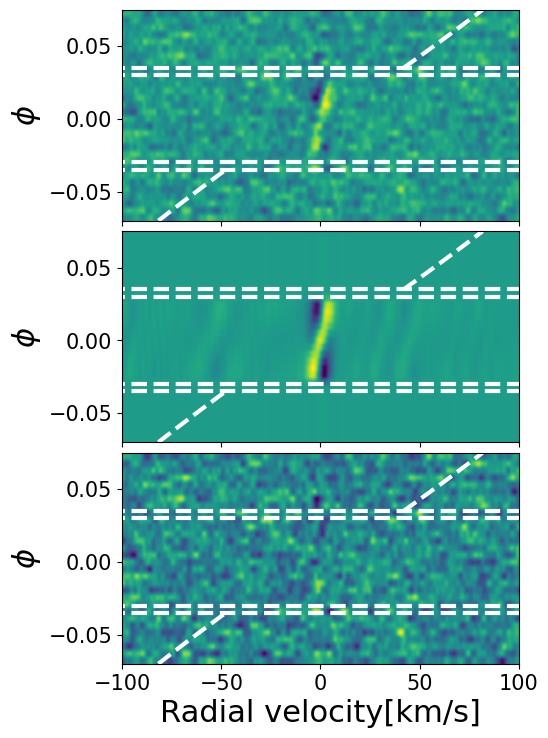}
    \caption{The combined CCF map for Ti at different orbital phase (\textit{top panel}). The map is combined with a step size of 0.005 in orbital phase between three nights. The white dashed lines mark the four contacts during the transit, the inclined white lines indicate the expected trace of atmospheric signal from the planet, and the bright yellow and dark stripe shows the RM+CLV signal. The \textit{middle panel} shows the predicted RM+CLV model map using \texttt{SME} and \texttt{petitRADTRANS}, which is used to eliminate the influence of the variation of stellar line profile during transit. The \textit{bottom panel} shows the residuals with RM+CLV effects corrected.}
    \label{fig:RC_model_and_ccfexample}
\end{figure}

\begin{figure} 
   \includegraphics[width=9cm, height=5.5cm]{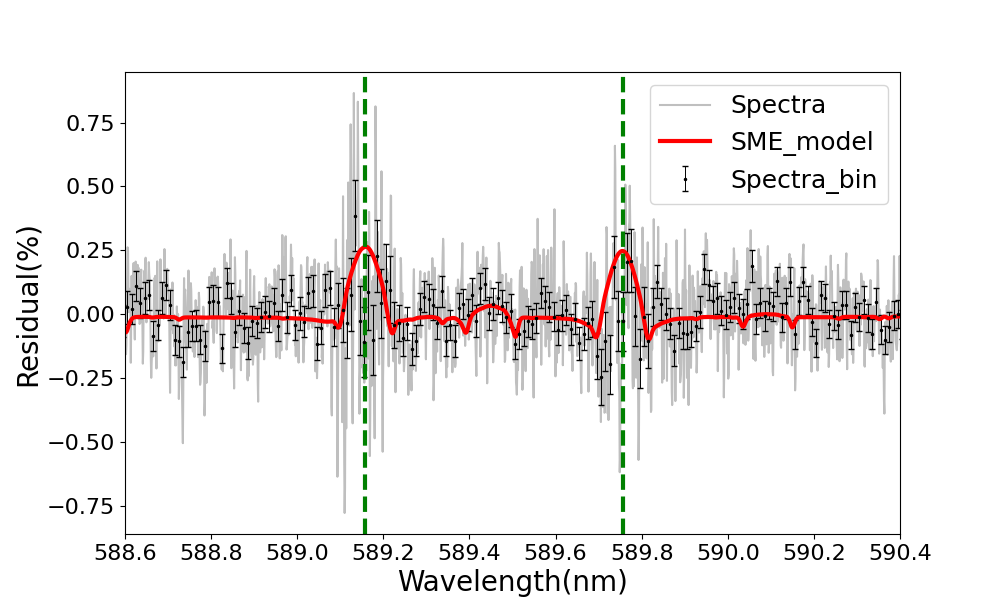}
   \centering
   \caption{Simulated CLV and RM effects overplotted in red on the transmission spectrum. The  transmission spectrum is shown in gray, and the spectrum binned by 0.1{\AA} is shown in black.}
   \label{clvrm}
    \end{figure}

\begin{figure*}
  \centering
  \includegraphics[width=17cm]{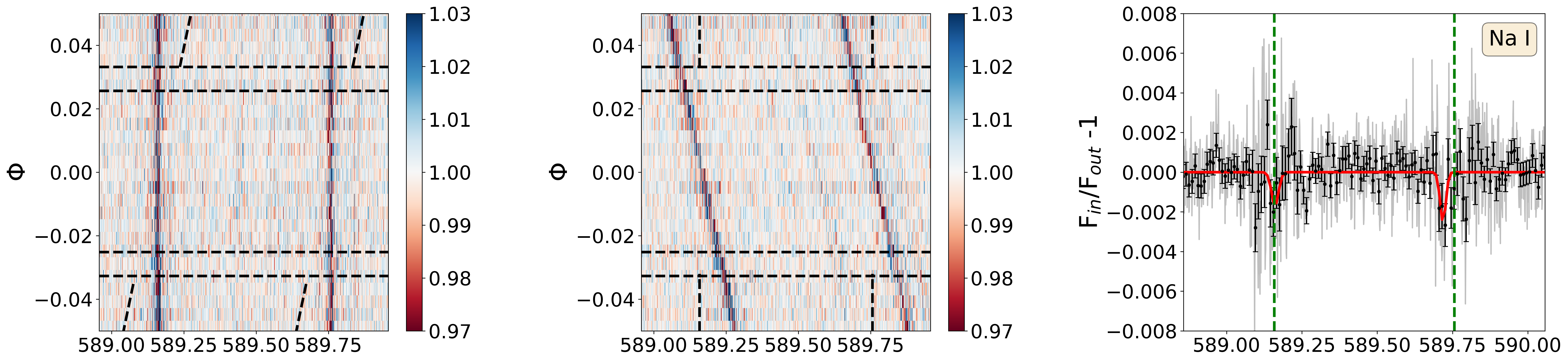}
  \includegraphics[width=17cm]{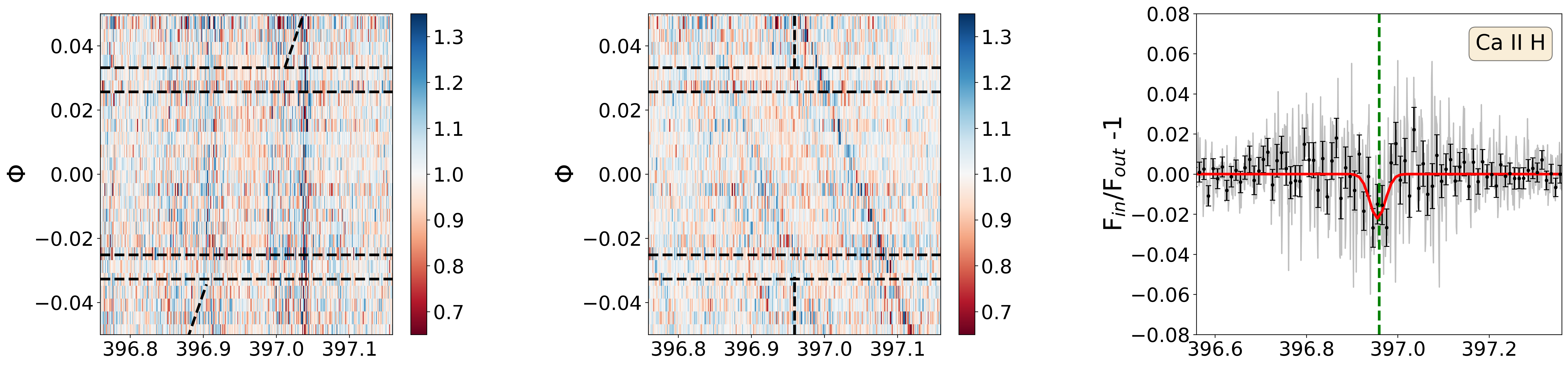}
  \includegraphics[width=17cm]{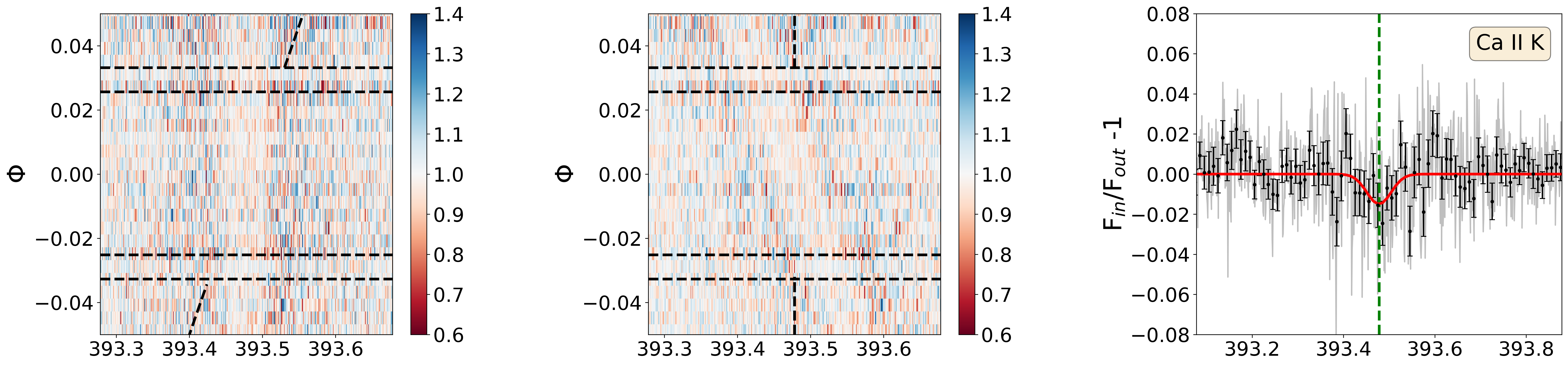}
  \includegraphics[width=17cm]{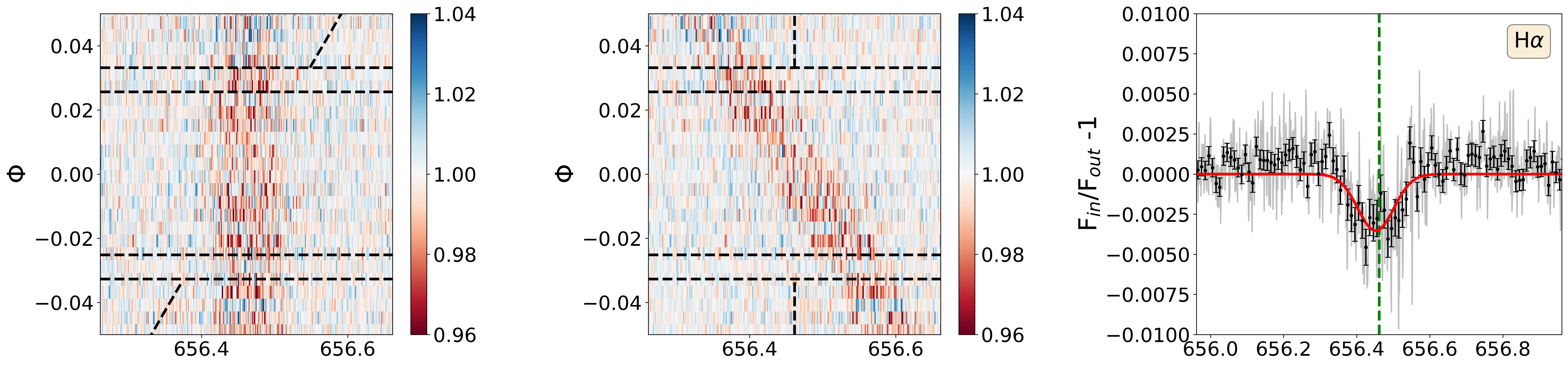}
  \includegraphics[width=17cm]{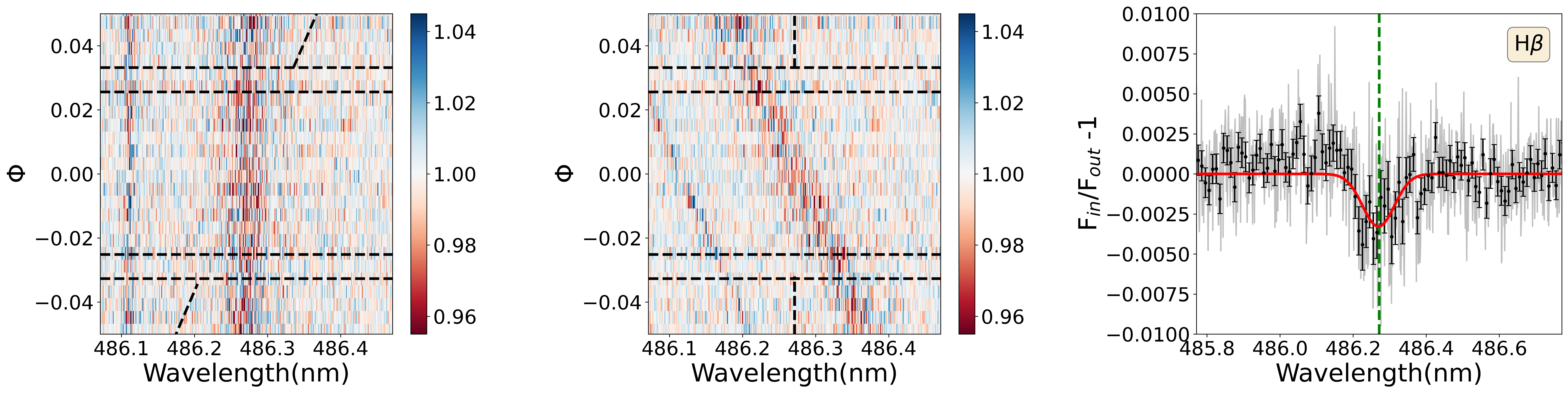}
  \caption{The phase-resolved 2D transmission spectra based on the combination of three night observations for Na double lines, \ion{Ca}{ii} H \& K, H$\alpha$ and H$\beta$ lines in the left panels. The horizontal black-dashed lines denote the beginning and end of the transit. The inclined black-dashed line presents the anticipated signal trajectory from the exoplanet atmosphere. The middle panels show the 2D transmission spectra in the planet rest frame (PRF) assuming $K_{\rm p}$= 191.94\,km\,s$^{-1}$. In the right panels, the combined transmission spectra in the PRF are shown in grey (original) and black (binned) with the optimal Gaussian fit represented in red. The dashed green vertical line represents the static position of each line at its vacuum wavelength. All spectra have been corrected for the CLV+RM effects.}
  \label{fig:atom line}
\end{figure*}

An illustrative example is presented in the upper panel of Figure~\ref{fig:RC_model_and_ccfexample} for the titanium (Ti) atom. The inclined white-dashed lines denote the anticipated trace of atmospheric absorption (if present) from the planet, while the alternating bright and dark structure is resulted from the CLV~\citep{Yanfei_2017} and the RM effect. These variations arise from the non-uniform brightness distribution across the stellar disk and the differing projected RVs at varying distances from the stellar center prior to the rotation of the host star. The upper panel depicts the observed CCF map, whereas the lower panel illustrates the model-predicted map, which is to be further elaborated upon in Section~\ref{sec:remove rm and clv effect}. 
%the bright slightly-inclined stripe
%%%%%%%%%%%%%%%%%%%%%%%%%%%%%%%%%%%%%%%%%%%%%%%%%%%%%
%%%%%%%%%%%%%%%%%%%%%%%%%%%%%%%%%%%%%%%%%%%%%%%%%%%%%

%-----------------------------------------------------    
%------------------------------------------------------

%-----------transmission-spectra-------------
%------------------------------------------------

%---------------------------------------------------%
 \begin{table*}
      \caption{Summary of the derived parameters of the atomic lines from the 3-night combined transmission spectrum}
         \label{tab:atomic_lines}
         %\centering
         \begin{tabular}{lccccc}
            \hline
            \hline
            \noalign{\smallskip}
            Species & $\lambda^1$ &h&$V_{\rm wind}$& FWHM&R$_{\lambda}$\\
                    & [nm]&[\%] &[km s$^{-1}$]&[km s$^{-1}$] & [$R_{\rm p}$]\\
            \noalign{\smallskip}
            \hline
            \ion{Na}{i}          &589.756/589.158 &0.21\,$\pm\,0.08$  & -8.97.\,$\pm\,2.25$&12.52\,$\pm\,5.32$& 1.06\,$\pm\,0.04$\\
            \ion{Na}{i} (masked) &   -        &0.30\,$\pm\,0.09$  & -10.24\,$\pm\,1.28$&8.83\,$\pm\,3.18$& 1.09\,$\pm\,0.05$\\
            \ion{Ca}{II} H       &396.959 &2.21\,$\pm\,0.63$  & -2.56\,$\pm\,4.18$&30.48\,$\pm\,9.97$&1.54\,$\pm\,0.25$\\
            \ion{Ca}{II} H (masked)  &- &1.70\,$\pm\,0.65$  & -2.58\,$\pm\,5.36$&29.26\,$\pm\,12.69$&1.43\,$\pm\,0.29$\\
            \ion{Ca}{II} K&393.478 &1.50\,$\pm\,0.57$  & -0.16\,$\pm\,9.07$&47.88\,$\pm\,21.80$&1.39\,$\pm\,0.25$\\
            \ion{Ca}{II} H (masked) &- &1.37\,$\pm\,0.57$  & 0.78\,$\pm\,9.75$&47.32\,$\pm\,23.47$&1.36\,$\pm\,0.26$\\
            H$\alpha$ &656.461 &0.35\,$\pm\,0.07$  & -4.16\,$\pm\,4.57$&52.85\,$\pm\,10.09$&1.10\,$\pm\,0.04$\\
            H$\alpha$ (masked)& - &0.08\,$\pm\,0.08$  & 4.38\,$\pm\,17.13$ &33.92$\pm\, 39.27$&1.02\,$\pm\,0.05$\\
            H$\beta$&486.271 &0.33\,$\pm\,0.08$  & -0.98\,$\pm\,6.97$&63.09\,$\pm\, 14.67$& 1.10\,$\pm\,0.05$\\
            H$\beta$ (masked)&- &0.13\,$\pm\,0.10$  & -16.09\,$\pm\,13.99$&37.81\,$\pm\, 30.89$& 1.04\,$\pm\,0.06$\\
            \noalign{\smallskip}
            \hline
            \noalign{\smallskip}
         \end{tabular}
         
         \textbf{Notes}:The line center wavelength is in vacuum, the depth of absorption line $h$, the Doppler shift of line center $V_{\rm wind}$, line width (FWHM) and the effective planetary radius, $R_{\lambda}$.
     \end{table*}
% %--------------------------------------------------%

   \begin{table}
   %\centering
   
      \caption{The source of H$\alpha$ absorption depth and XUV data.}
         \label{Tab:Ha_and_XUV} 
           \begin{tabular}{ccc}
            \hline
            \hline
            Planet &\qquad H$\alpha$ data source & \qquad XUV data source \\
            \hline
            KELT-9\,b&\qquad1& \qquad 7  \\ 
            HD 209458\,b& \qquad 2&\qquad 8  \\ 
            HD 189733\,b&\qquad 3&\qquad 8\\ 
            WASP-77A\,b& \qquad This paper & \qquad 8  \\ 
            WASP-52\,b&\qquad 4\,& \qquad 7  \\   
            WASP-121\,b&\qquad 5&\qquad 9\\
            WASP-19\,b&\qquad 6&\qquad 9\\
            \hline
            \noalign{\smallskip}
         \end{tabular}
         
        \textbf{References}:$(1)$~\citet{Wyttenbach_2020}, $(2)$~\citet{Casasayas_2021}, $(3)$~\citet{Cauley_2015}, $(4)$~\citet{Chen_2020}, $(5)$~\cite{Maguire_2023}, $(6)$~\citet{Sedaghati_2021}, $(7)$~\citet{Nortmann_2018}, $(8)$~\citet{Salz_2016}, $(9)$~\citet{Foster_2022}.
   \end{table}
%%%%%%%%%%%%%%%%%%%%%%%%%%%%%%%%%%%%%%%%%%%%%%%%%%

\section{RESULTS AND DISCUSSION}
In the preceding sections, we detailed the observation of three transits of the hot Jupiter WASP-77Ab utilizing the ESPRESSO spectrograph. A total of 157 spectra were collected, comprising 67 spectra acquired during the transits and 90 spectra obtained outside of transit periods. Each spectrum underwent correction for telluric contamination through the application of  \texttt{Molecfit}. Consequently, the master out-of-transit spectra were constructed, and the transmission spectra were generated by integrating data from the three observational nights.

\begin{figure*}
  \centering
  \includegraphics[width=17cm]{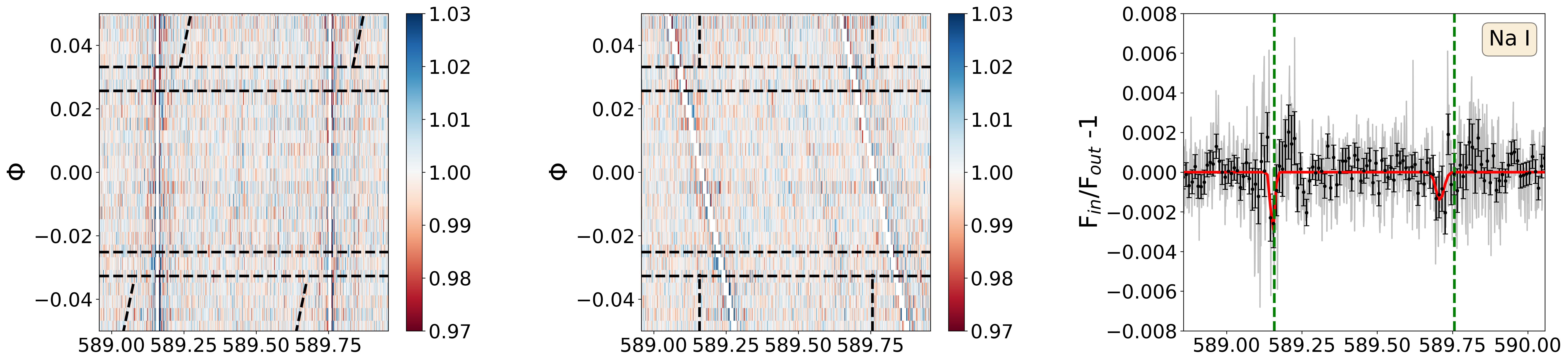}
  \includegraphics[width=17cm]{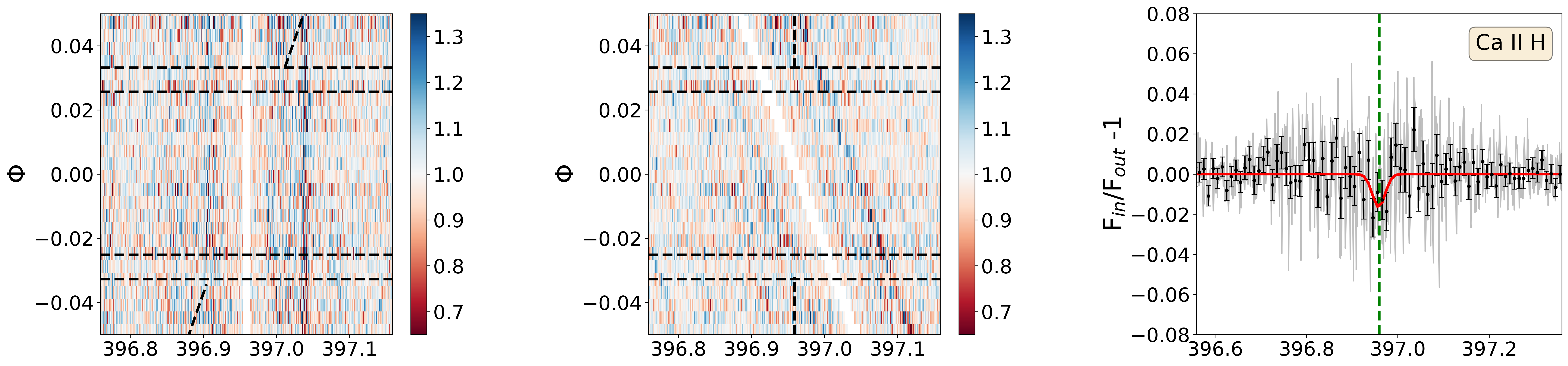}
  \includegraphics[width=17cm]{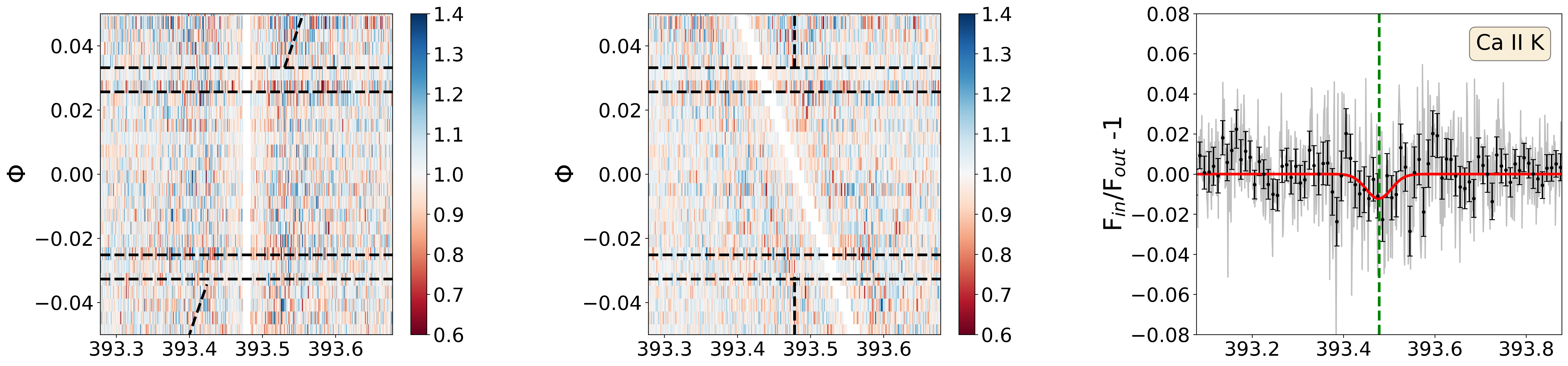}
  \includegraphics[width=17cm]{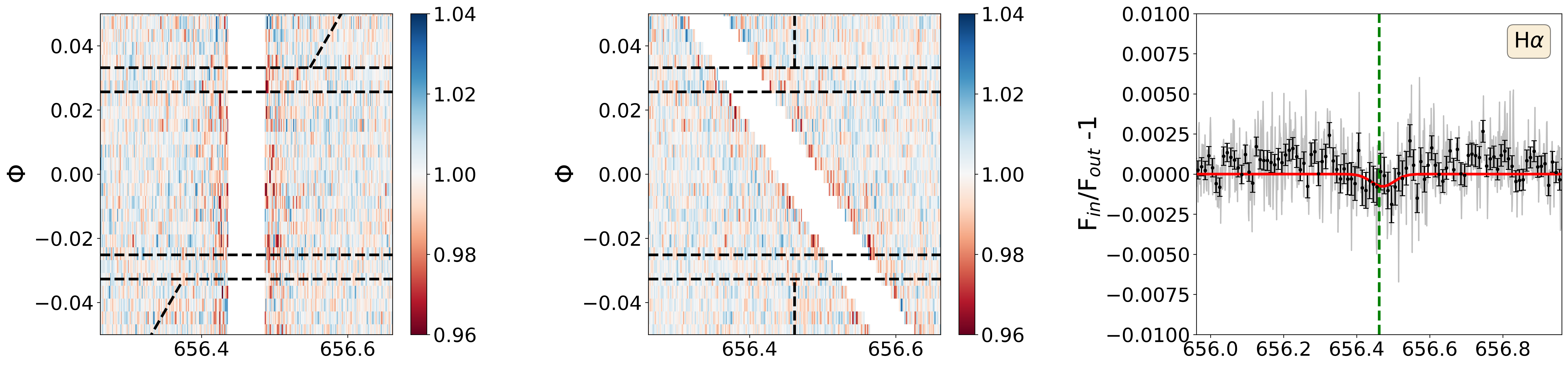}
  \includegraphics[width=17cm]{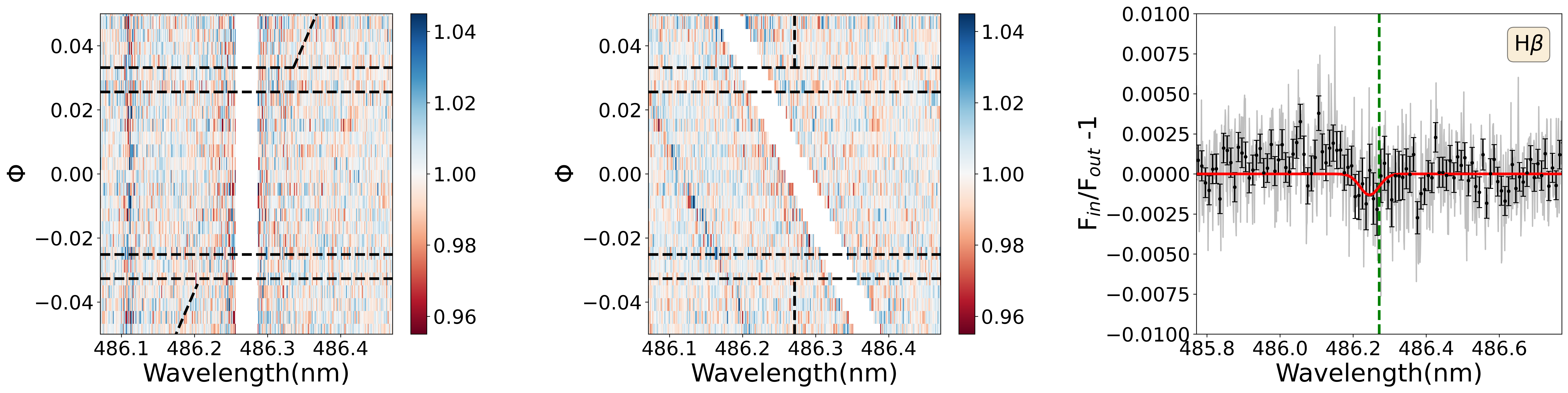}
  \caption{Same as Fig~\ref{fig:atom line}: but we mask the line cores with widths of 0.1{\AA} for the \ion{Na}{i} \& \ion{Ca}{ii} doublet, 0.3{\AA} for H$\beta$ and 0.5{\AA} for H$\alpha$, respectively, where the SNRs are low and the corrected RM+CLV residuals may still exist.}
  \label{fig:mask atom line}
\end{figure*}

We report, for the first time, the detection of additional absorption of H$\alpha$ and H$\beta$ with significance levels of $\sim5\sigma$ and 4$\sigma$, respectively, in the atmosphere of WASP-77Ab. In addition, we report a tentative detection of the \ion{Ca}{II} H \& K lines as well with significance levels of $\sim$3.5$\sigma$ and 2.6$\sigma$, respectively, as detailed in Table~\ref{tab:atomic_lines} and illustrated in Fig.~\ref{fig:atom line}. It is important to note that residual structures remain in the 2D transmission spectra and the core lines are considerably influenced by stellar residuals, which may arise from the imperfect RM+CLV correction, or from noise associated with low SNRs in the stellar line cores. To address this issue, we followed the procedures outlined in \citet{Jiang_2023a} and applied masks with widths of 0.1{\AA} to the \ion{Na}{i} and \ion{Ca}{ii} doublets, 0.3{\AA} to H$\beta$ and 0.5{\AA} to H$\alpha$, respectively. This approach aimed to minimize potential stellar contamination and to avoid regions with low SNRs. By doing so, we were able to identify any residual signals that may originate from the planet. As illustrated in Fig.~\ref{fig:mask atom line}, all absorption features observed in Fig.~\ref{fig:atom line} still persist, albeit at slightly reduced strength, implying that the observed absorptions may partially originate from the planetary atmosphere. However, we cannot rule out other possibilities, such as residual stellar effects that may persist due to the imperfect correction of RM+CLV effects.

We also note the majority of the detected lines exhibit a blueshift relative to their rest-frame positions, as listed in Table~\ref{tab:atomic_lines}, which is commonly inferred in the transmission spectra of HJs, and is widely attributed to the influence of planetary winds in the terminator region of tidally-locked exoplanets \citep{Casasayas-Barris_2019,Tabernero_2021,Kesseli_2022}. Furthermore, we conducted searches for other atomic species, including Cr, Fe, Ti, V, and Y, as well as molecular species such as TiO, VO, HCN, H$_2$O, CaH, and FeH using the CCF technique. However, these searches yielded no positive results for either atomic or molecular detection, as illustrated in Figures~\ref{fig:ccf_result_petit_1} and \ref{fig:ccf_result_petit_2}, respectively. These molecules are recognized as significant indicators of thermal inversion \citep{Hubeny_2003,Knutson_2008,Evans_2016}. Despite the non-detection of these species, we cannot dismiss the potential presence of other species that may contribute to thermal inversion in the atmosphere of WASP-77Ab.

Previous studies have identified the presence of H$_2$O and CO in the dayside atmosphere, as reported by \citet{Line_2021, Mansfield_2022, August_2023}. More recently, \citet{Smith2024} reported a $>10\sigma$ detection of H$_2$O  and a $5\sigma$ detection of CO, respectively. In the CCF map (their Fig.~4), the planet signal trails persist almost uniformly until the end of the observed pre-eclipse phase at $\sim$0.32, which is near the quadrature position. This observation implies that the day side portion of the terminator region likely retains a substantial concentration of H$_2$O vapour. Although our ESPRESSO transmission spectra do not reveal any CCF signal of H$_2$O, it remains uncertain whether the terminator region of WASP-77Ab contains water vapour. This is because the ESPRESSO passband of $380-788$\,nm is much less sensitive to H$_2$O compared to the NIR and IR instruments employed by \citet{Smith2024}. Obtaining transmission spectra in the NIR and IR bands would offer a comprehensive view of the longitudinal distribution of H$_2$O in the atmosphere of WASP-77Ab.

H$\alpha$ serves as a highly sensitive indicator for  exploring the structure and heating processes of the upper atmosphere of exoplanets that experience significant irradiation. The excitation and de-excitation processes involving H atoms are closely linked to local particle densities, temperatures, and the radiation fields in the atmospheres of these exoplanets atmosphere~\citep{Huang_2017}. Despite the detection of other molecular species, such as H$_2$O, in several tens of hot Jupiters, H$\alpha$ has been reliably identified in just about 10 ultra-hot Jupiters (UHJs, with temperatures exceeding about 2000\,K), including WASP-121b~\citep{Maguire_2023} and KELT-9b~\citep{Yan_2018,Wyttenbach_2020}. The significantly low detection rate of H$\alpha$ in classical HJs is posited to be due to the inadequacy of XUV stellar radiation in these systems to promote the escape of neutral hydrogen and the formation of a substantial, low-density, and elevated-temperature hydrogen halo.

Given the variability of XUV flux, this particular population of exoplanets warrants further investigation, as it may elucidate the mechanisms by which planetary atmospheres are heated and cooled under varying XUV conditions. The sources for both H$\alpha$ absorption depth and XUV data for each planet are detailed in Table~\ref{Tab:Ha_and_XUV}. It is noteworthy that several planets with H$\alpha$ absorption measured, such as WASP-33\,b, 76\,b, 189\,b, are excluded from this table due to the unavailability of XUV flux data from their respective host stars. To establish a meaningful observational relationship between H$\alpha$ absorption and incident XUV flux, the current sample size is limited. Expanding the sample to include more planets with both H$\alpha$ and XUV flux measurements will significantly enhance our understanding of the heating and cooling processes occurring in the upper atmospheres of HJs and UHJs.

%-----------masked-transmission-spectra----------
%-----------------------------------------------

%%%%%%%%%%%%%%%%%%%%%%%%%%%%%%%%%%%%%%%%%%%%%%%%%%
\section*{Acknowledgements}
This research is supported by the National Natural Science Foundation of China grants No.11988101, 42075123, 42005098, 62127901, 12273055, the National Key R\&D Program of China No.~SQ2024YFA1600043, the China Manned Space Project with NO. CMS-CSST-2021-A11 and the Pre-research project on Civil Aerospace Technologies No. D010301 funded by China National Space Administration (CNSA). MZ are supported by the Chinese Academy of Sciences (CAS), through a grant to the CAS South America Center for Astronomy (CASSACA) in Santiago, Chile.
\section*{Data availability}
The observations underlying this analysis are publicly available in
the ESO Science Archive Facility under program name 0108.C-0148.
Other data will be shared on reasonable request to the corresponding
author.


\bibliographystyle{mnras}
\bibliography{citepaper} % if your bibtex file is called example.bib

%\begin{thebibliography}{99}
%\end{thebibliography}

%%%%%%%%%%%%%%%%% APPENDICES %%%%%%%%%%%%%%%%%%%%%

\appendix

\section{Some extra material}

%%%%%%%%%%%%%%%%%%%%%%%%%%%%%%%%%%%%%%%%%%%%%%%%
\begin{figure*}
   \includegraphics[width=13cm]{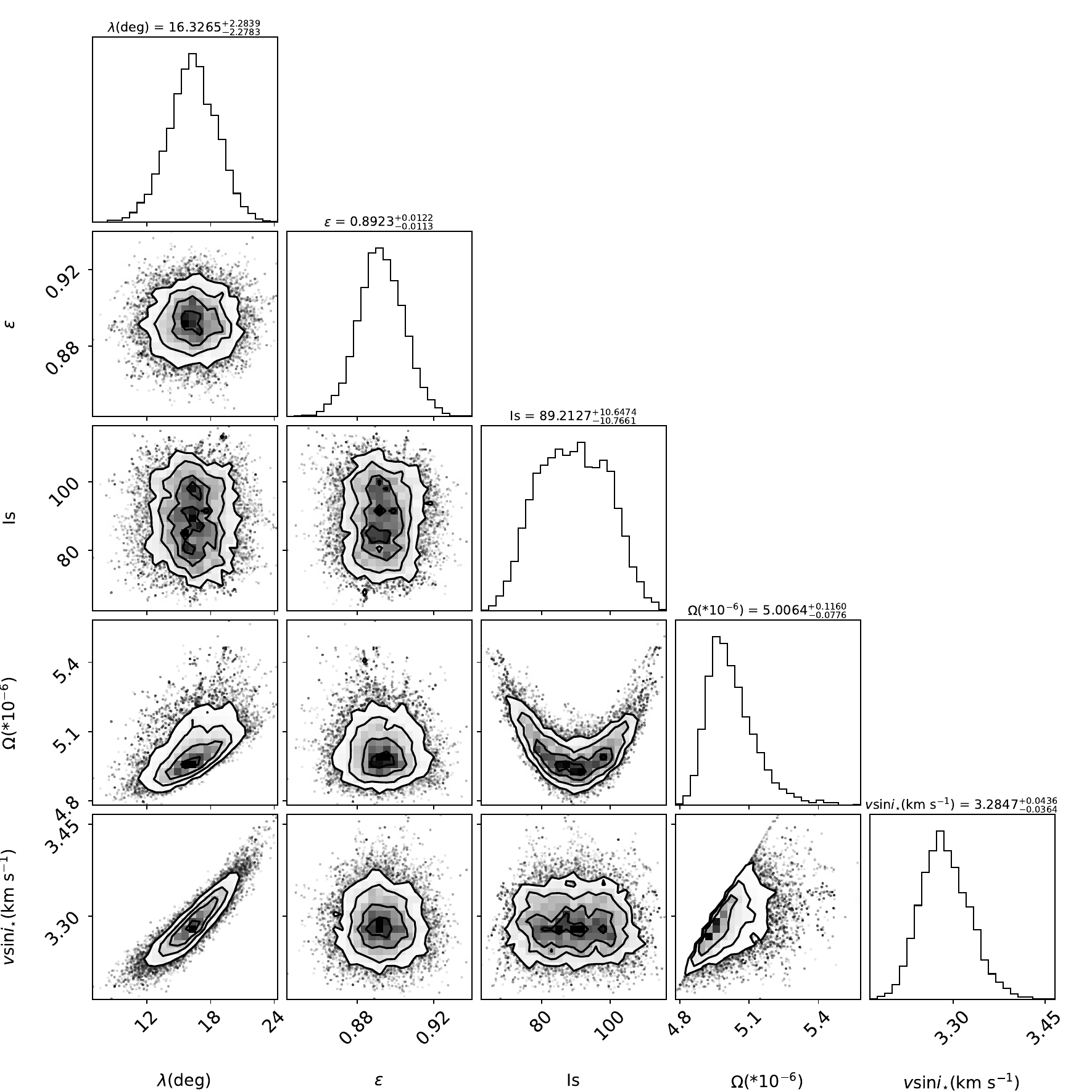}
   \centering
   \caption{Posterior distribution of the best-fit RM model using PyAstronomy.}
   \label{fig:RM_fit_corner}
    \end{figure*}
%%%%%%%%%%%%%%%%%%%%%%%%%%%%%%%%%%%%%%%%%%%%%%%%

%%%%%%%%%%%%%%%%%%%%%%%%%%%%%%%%%%%%%%%%%%%%%%%%
\begin{figure*}
    \centering
    \subfigure{
    \begin{minipage}[t]{1\linewidth}
    \centering
    \includegraphics[width=17cm]{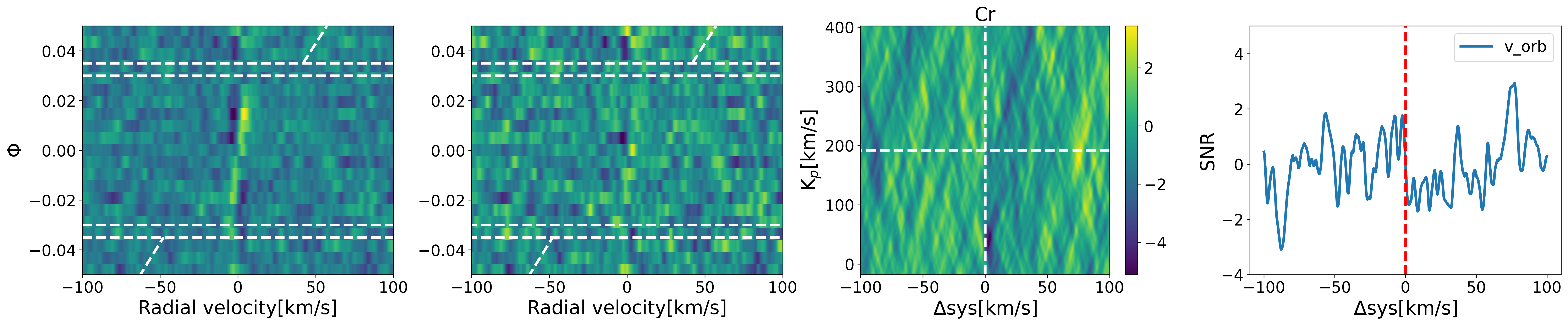}  
    \end{minipage}
    }
    
    \subfigure{
    \begin{minipage}[t]{1\linewidth}
    \centering
    \includegraphics[width=17cm]{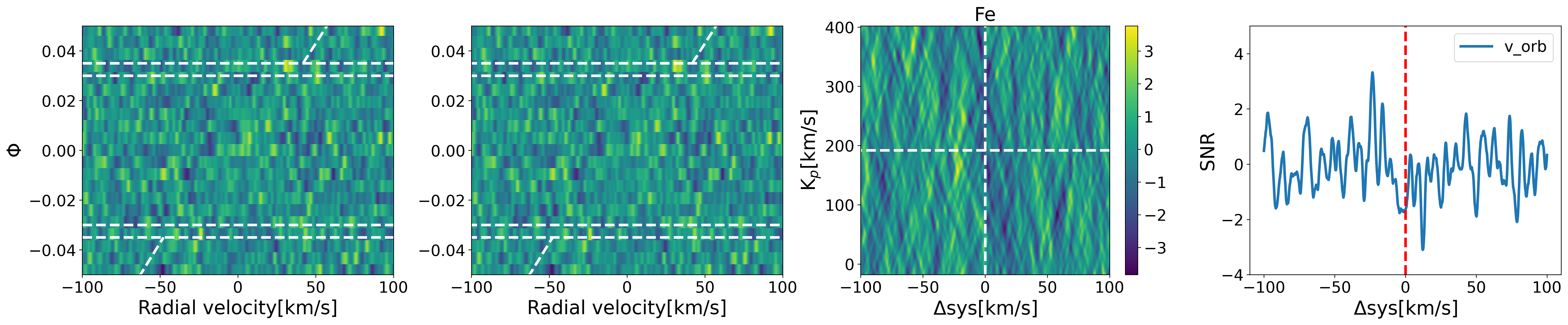}  
    \end{minipage}
    }
    
    \subfigure{
    \begin{minipage}[t]{1\linewidth}
    \centering
    \includegraphics[width=17cm]{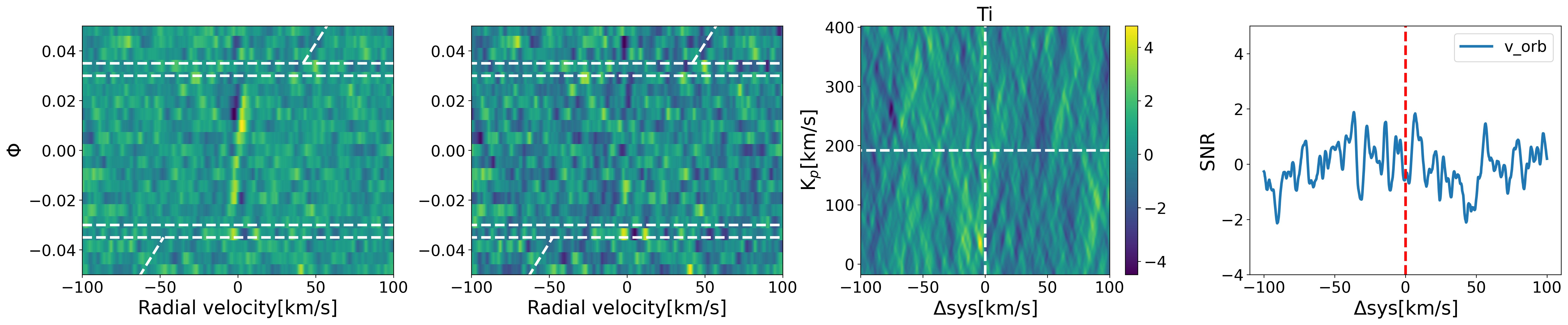}  
    \end{minipage}
    }

    \subfigure{
    \begin{minipage}[t]{1\linewidth}
    \centering
    \includegraphics[width=17cm]{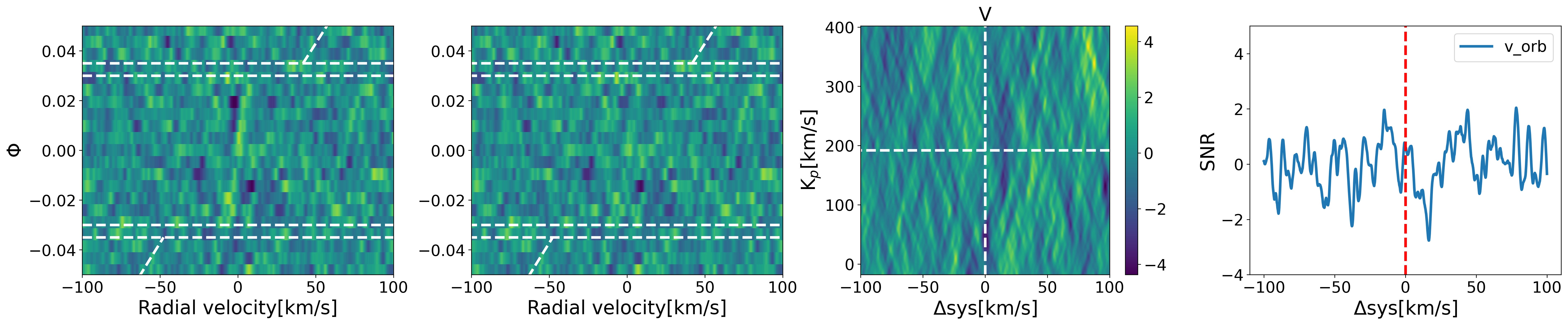}  
    \end{minipage}
    }
    
    \subfigure{
    \begin{minipage}[t]{1\linewidth}
    \centering
    \includegraphics[width=17cm]{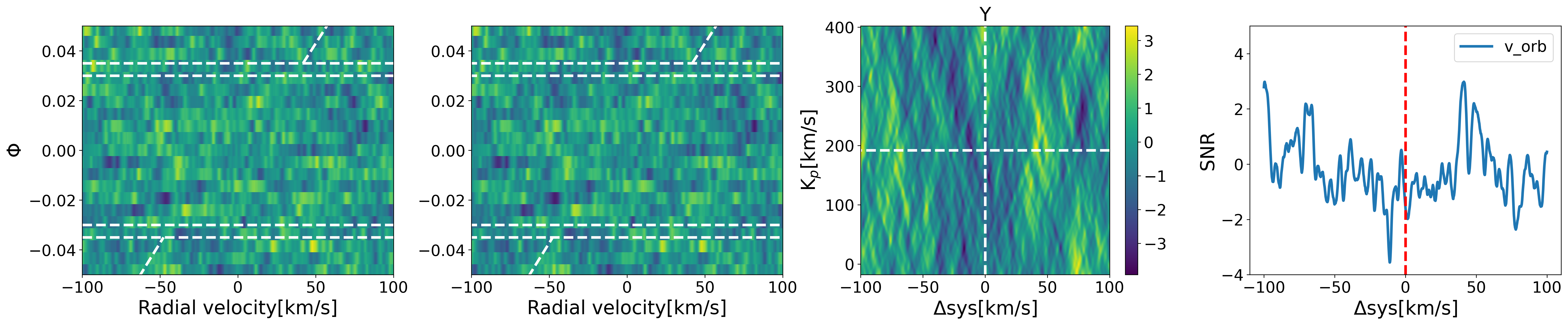}  
    \end{minipage}
    }
    
\caption{\emph{First panels}: The 2D CCF maps of Cr, Fe, Ti, V and Y with CLV+RM effects uncorrected using \texttt{pRT} opacity. The white dotted lines mark the beginning and ending positions of the transit and the inclined white lines indicate the expected trace of signal from the planet if exists. \emph{Second panels}: Same as \emph{the first panels} but with CLV+RM effects corrected. \emph{Third panels}: the $K_{\rm p}$-$\Delta V_{\rm sys}$ maps in the range of $-20\sim400$\, km\,s$^{-1}$. In each panel, the planet signal is expected to appear around the intersection of the two white dotted lines. \emph{Fourth panels}: the SNR plots at the expected $K_{\rm p}$ position in blue.}
    \label{fig:ccf_result_petit_1}
\end{figure*}
%%%%%%%%%%%%%%%%%%%%%%%%%%%%%%%%%%%%%%%%%%%%%%%%
%%%%%%%%%%%%%%%%%%%%%%%%%%%%%%%%%%%%%%%%%%%%%%%%
\begin{figure*}
    \centering
    \subfigure{
    \begin{minipage}[t]{1\linewidth}
    \centering
    \includegraphics[width=17cm]{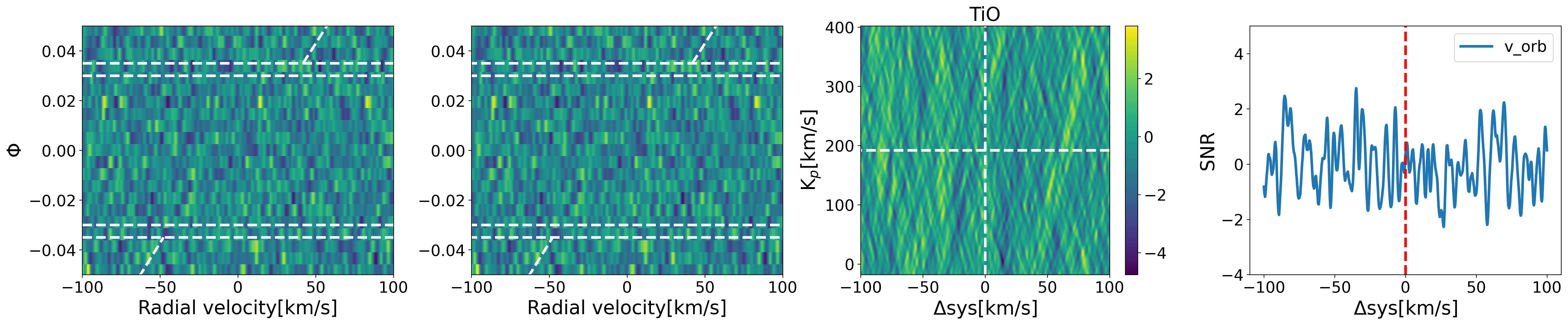}  
    \end{minipage}
    }
  
    \subfigure{
    \begin{minipage}[t]{1\linewidth}
    \centering
    \includegraphics[width=17cm]{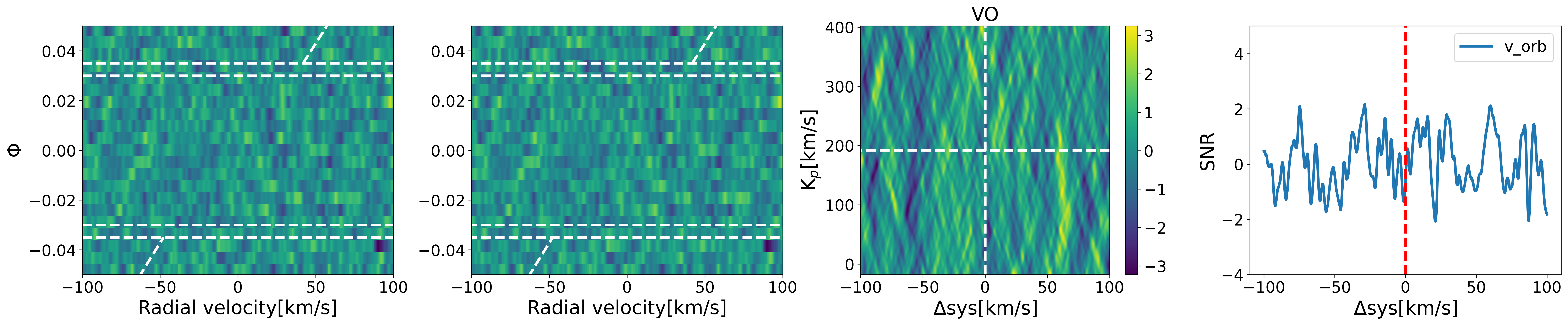}  
    \end{minipage}
    }
    
    \subfigure{
    \begin{minipage}[t]{1\linewidth}
    \centering
    \includegraphics[width=17cm]{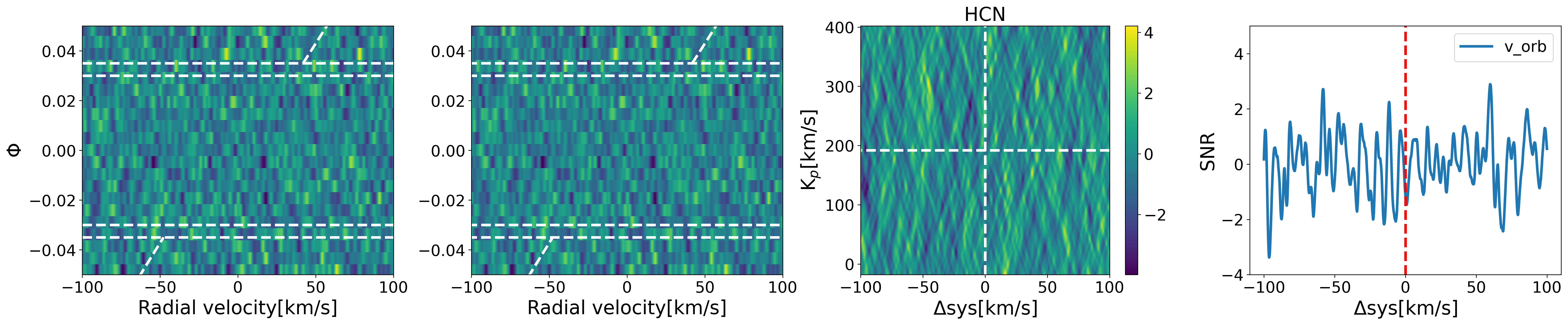}  
    \end{minipage}
    }
    
    \subfigure{
    \begin{minipage}[t]{1\linewidth}
    \centering
    \includegraphics[width=17cm]{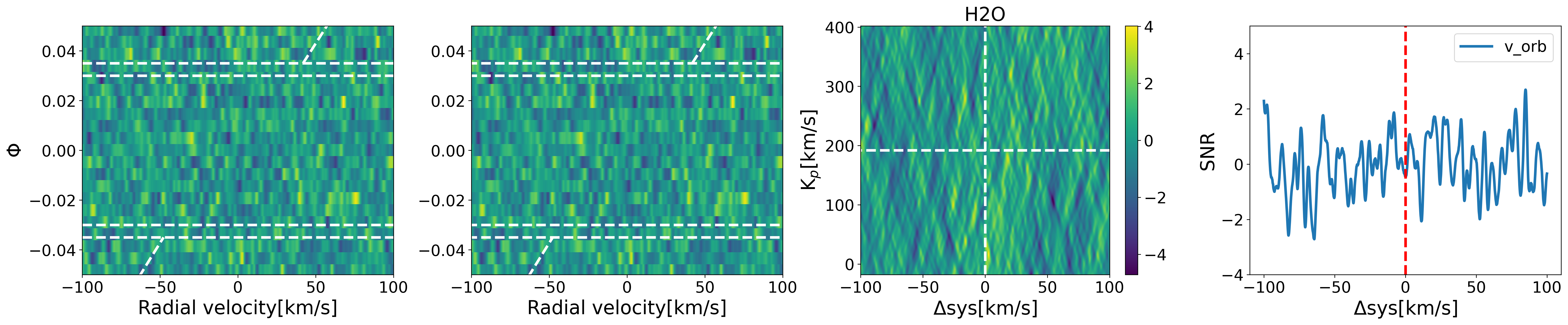}  
    \end{minipage}
    }
    
    \subfigure{
    \begin{minipage}[t]{1\linewidth}
    \centering
    \includegraphics[width=17cm]{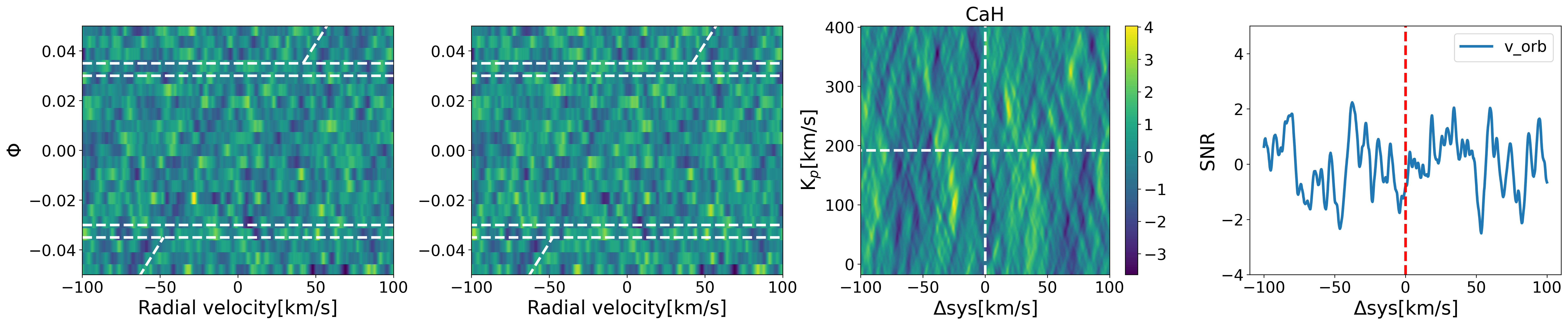}  
    \end{minipage}
    }
    
    \subfigure{
    \begin{minipage}[t]{1\linewidth}
    \centering
    \includegraphics[width=17cm]{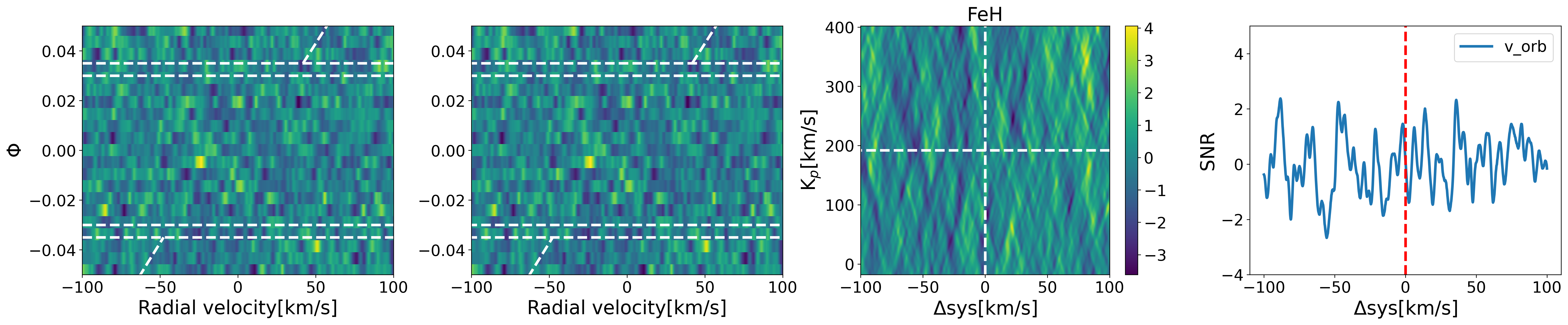}  
    \end{minipage}
    }

\caption{Same as Fig~\ref{fig:ccf_result_petit_1}: but for TiO, VO, HCN, H2O, CaH and FeH}
    \label{fig:ccf_result_petit_2}
\end{figure*}
%%%%%%%%%%%%%%%%%%%%%%%%%%%%%%%%%%%%%%%%%%%%%%%

% Don't change these lines
\bsp	% typesetting comment
\label{lastpage}
\end{document}